\documentstyle[preprint,aps,epsf]{revtex}       
\global\let\epsfloaded=Y 
\begin{document}
\pagestyle{empty}                                
\preprint{
\font\fortssbx=cmssbx10 scaled \magstep2
\hbox to \hsize{
\hfill $
\vtop{
 \hbox{ }}$
}
}
\draft
\vfill
\title{ Radiative Decay of $\Upsilon$ into a
Scalar Glueball}
\vfill
\author{
X.-G. He$^{1,2}$, H.-Y. Jin$^1$  and J. P. Ma$^{1,3}$}

\address{
$^1$Institute of Theoretical Physics, 
Zhejiang University, Zhejiang\\
$^2$Department of Physics, National Taiwan University, Taipei\\
$^3$Institute of Theoretical Physics , Academia
Sinica, Beijing }

%
%
\vfill
\maketitle
\begin{abstract}
We study the radiative decay of $\Upsilon$ into a scalar glueball
$\Upsilon \to \gamma G_s$ using QCD factorization. We find that
for this process the non-perturbative effects can be factorized
into a matrix element well defined in non-relativistic QCD (NRQCD)
and the gluon distribution amplitude. The same NRQCD matrix
element appears also in leptonic decay of $\Upsilon$ and therefore
can be determined from data.  In the asymptotic limit the gluon
distribution amplitude is known up to a normalization
constant. Using a QCD sum-rule calculation for the normalization
constant, we obtain $Br(\Upsilon \to \gamma G_s)$ to be in the
range $(1\sim 2)\times 10^{-3}$. 
We also discuss some of the implications for $\Upsilon \to \gamma 
f_i$ decays. 
Near future data from CLEO-III
can provide crucial information about scalar glueball properties.

\end{abstract}
%
%
%
%
\pagestyle{plain}

\noindent
{\bf 1. Introduction}

The existence of glueballs are natural predictions of QCD. Some of
the low lying
states are $0^{++}$, $0^{-+}$, $1^{+-}$ and 
$2^{++}$ with the lowest
mass eigenstate to be $0^{++}$ in the range of  $1.5\sim 1.7$ GeV
from theoretical calculations\cite{2}.
There are indications that $f_0(1370)$, 
$f_0(1500)$ and $f_0(1710)$ contain
substantial scalar glueball content.
For the search of glueballs,
decays of quarkonia are well suited processes
because the decays are mediated by gluons. Among these
decays, two-body radiative decays are ideal places
to study this subject, because there is no
complication of interactions between light hadrons.
Radiative decays of $\Upsilon$ have been
studied before, in particularly by CLEO\cite{3,4} recently.
With the current data sample, 
there are already several observations of
radiative decay of $\Upsilon$ into mesons. Among them
only a few with good precisions, such as
the decay $\Upsilon\to\gamma f_2(1270)$, 
$\Upsilon \to \gamma f_0(1710)
\to \gamma K \bar K)$, while the others have large errors\cite{4}.
About $4 ~{\rm fb}^{-1}$ $b{\bar b}$ resonance data are
planned to be taken at CLEO-III in 
the year prior to conversion to low
energy operation (CLEO-C)\cite{1}. 
This will produce the largest data
sample of $\Upsilon$ in the world.
More radiative decay modes of 
$\Upsilon$ may be observed. Combining
experimental data in the near future and
theoretical results
glueball properties can be studied in details.

In this paper
we carry out a theoretical study of
the radiative decay of $\Upsilon$ into a scalar glueball by using
QCD factorization.
We find that the non-perturbative
effects can be factorized into a matrix element well defined in
non-relativistic QCD (NRQCD), and the
gluon distribution amplitude.
The NRQCD matrix element can be determined from leptonic
$\Upsilon$ decays. The asymptotic form of the gluon distribution
amplitude is 
known in QCD up to a normalization constant. Using a
QCD sum rule calculation for this constant, the branching ratio
$Br(\Upsilon \to \gamma G_s)$ is predicted to be in the range of
$(1\sim 2)\times 10^{-3}$.
Combining this result with 
experimental data, we find that none of the
candidate scalar glueballs 
$f_0(1500)$ and $f_0(1710)$ can be a pure
glueball. Existing information on glueball mixing allow us
to predict the branching ratios for several radiative decays,
such as $\Upsilon \to \gamma f_0(1370, 1500,1710) \to
\gamma K \bar K (\pi\pi)$. A mixing pattern suggested in the
literature is shown to be in conflict with data.
Near future experimental data from CLEO-III will provide
crucial information about scalar glueball properties.
\\

\noindent
{\bf 2. QCD factorization of $\Upsilon \to \gamma G_s$}

\par
It is known that properties 
of $\Upsilon$ can be well described
with non-relativistic
QCD(NRQCD)\cite{5}. The decay of $\Upsilon \to \gamma G_s$
can be thought of as a
free $b\bar b$ quark pair first
freed from $\Upsilon$ with a 
probability which is characterized
by matrix elements defined in NRQCD,
this pair of quarks decays into a photon and gluons,
and then the gluons subsequently converted
into a scalar glueball. 
In the heavy quark limit $m_b\to\infty$,
the glueball has a
large momentum, this allows a twist-expansion 
to describe the conversion.
Also, the gluons
are hard and perturbative 
QCD can be applied for the decay of the
$b\bar b$ pair into
a photon and gluons. 
This implies that the decay width can be factorized.
In the real world,
the b-quark mass is 5 GeV and
a scalar glueball has a mass around 1.5 GeV as suggested by
lattice QCD simulations\cite{3}. This
may lead to
a question if the twist expansion is applicable.
For the radiative decay of $\Upsilon$, the glueball has a
momentum of order of $m_b$. The
twist expansion means a 
collinear expansion of momenta of gluons
in the glueball,  components of these momenta have the
order of $({\cal O}(k^+),
{\cal O}((k^-),{\cal O}(\Lambda_{QCD}),
{\cal O}(\Lambda_{QCD}))$, 
where $k$ is the momentum of the glueball.
Here we used
the light-cone 
coordinate system. Hence the expansion parameters are
\begin{equation}
\frac{k^-}{k^+}=\frac{m^2_G}{M_{\Upsilon}^2}\sim 0.02, \ \
  \frac {\Lambda_{QCD}}{k^+}\sim  0.1,
\end{equation}
where $m_G$ is the mass of the glueball and
we have taken 
$\Lambda_{QCD}\approx 500$MeV. In the above estimation
we have used the fact that the probability
for the conversion of gluons into a glueball is
suppressed if the ``+'' component
of the momentum of a gluon is very small. 
We see that the relevant expansion
parameters are small, therefore twist expansion
is expected to be a good one. 
We note that the same approximation may not
be applied to $J/\psi$ system because 
in this case the relevant expansion
parameters are not small.
We now provide some details of the calculations.
\par

The leading Feynmann diagrams for
$\Upsilon \to \gamma G_s$ are from $b\bar b$ annihilation
into two gluons and a photon.
The basic formalism for such calculations have been developed
in Ref.\cite{6} and 
has been used in the case of 
$\Upsilon \to \gamma \eta (\eta')$ to obtained
consistent result with 
experimental data\cite{12}. 
With appropriate modifications we can obtain the
S-matrix for $\Upsilon \to \gamma G_s$ decay. It is given by

\begin{eqnarray}
\langle\gamma G_s|S|\Upsilon\rangle 
=&-& i{1\over 2} e Q_b g^2_s \epsilon^*_\rho
\int d^4x d^4y d^4z d^4x_1 d^4y_1 
e^{iq\cdot z} \langle G_s| G^a_{\mu}(x)
G^a_\nu(y)|0\rangle \nonumber\\
&&\langle 0\vert \bar b_j(x_1) b_i(y_1)
\vert\Upsilon\rangle 
\cdot M^{\mu\nu\rho,ab}_{ij}(x,y,x_1,y_1,z),
\end{eqnarray}
where
$M^{\mu\nu\rho,ab}_{ij}$ 
is a known function from evaluation of the
Feynmann diagrams, $i$ and $j$ stand for 
Dirac- and color indices, $a$ and $b$
is the color indices of gluon. $\epsilon^*$ 
is the polarization vector of the
photon and $Q_b=-1/3$ is the $b$ quark 
electric charge. Since $b$ quark is heavy
and moves with small velocity $v$, one 
can expand the Dirac fields in NRQCD fields:

\begin{equation}
\langle 0\vert\bar b_j(x) b_i(y)\vert\Upsilon\rangle 
= -{1\over 6} (P_+\gamma^l P_-)_{ij}
\langle 0|\chi^\dagger \sigma^l 
\phi|\Upsilon\rangle e^{-ip\cdot (x+y)} + O(v^2),
\end{equation}
where $\chi^\dagger (\psi)$ 
is the NRQCD field for $\bar b(b)$ quark and
$P_\pm = (1\pm \gamma^0)/2$. 
The $b$ is almost at rest, then $p_\mu = (m_b, 0,0,0)$
with $m_b$ being the $b$ quark pole mass.

From the above we obtain the decay 
amplitude for $\Upsilon \to \gamma G_s$ as

\begin{equation}
{\cal T} = \frac{eQ_b g_s^2}{6}\langle 0
 \vert \chi^\dagger {\bf \epsilon} 
\cdot\sigma\psi\vert\Upsilon\rangle
  \int_0^1 dz \frac{1}{z(1-z)} {\cal F}_s(z),
\end{equation}
the decay width then reads:
\begin{equation}
\Gamma = \frac{2}{9m_b^4} \pi^2 Q_b^2 \alpha\alpha_s^2
  \langle\Upsilon\vert O_1(^3S_1)\vert\Upsilon\rangle
  \left\vert \int_0^1 dz 
\frac{1}{z(1-z)} {\cal F}_s(z)\right\vert^2.
\end{equation}
In the above ${\cal F}_s$ is the gluon
distribution amplitude of $G_s$ and is given by

\begin{equation}
{\cal F}_s (z)
= \frac{1}{2\pi k^+} \int dx^-e^{-iz k^+x^-}
\langle G_s (k) \vert G^{a,+\mu}(x^-) 
{G^{a,+}}_\mu (0) \vert 0\rangle.
\label{fs}
\end{equation}
Here we have used a gauge with 
$G^+ = 0$ such that the gauge link
between the field strength operators vanish.
This distribution characterizes basically how two
gluons are converted into $G_s$, where one
of the two gluons has the momentum $(zk^+,0,{\bf O_T})$.

In the above equations, the
matrix element
$\langle\Upsilon\vert 
O_1(^3S_1)\vert\Upsilon\rangle$ is defined
in NRQCD contains
the bound state effect 
of b-quarks in $\Upsilon$\cite{5} and can
be extracted from leptonic $\Upsilon \to l^+l^-$ decay.
A prediction can be made for $\Upsilon \to \gamma G_s$
if the distribution amplitude is known.
\par
The distribution amplitude can be written as
\begin{equation}
{\cal F}_s(z) = 
f_s f(z),\ \ \ {\rm with}\ \int_0^1 dz f(z) =1.
\end{equation}
where $f(z)$ is a 
dimensionless function and its asymptotic form
is:
\begin{equation}
f(z) = 30z^2(1-z)^2.
\end{equation}
With the asymptotic form in Eq.(6) we have:
\begin{equation}
R_s= \frac{\Gamma(\Upsilon\to\gamma 
+ G_s)}{\Gamma(\Upsilon\to \ell^+\ell^-)}
  = \frac{25\pi\alpha_s^2}{3\alpha} 
\cdot\frac{\vert f_s\vert^2}{m_b^2}.
\label{R}
\end{equation}
In the above we have used the fact 
that both $\Upsilon \to \gamma G_s$ and
$\Upsilon \to l^+l^-$ are proportional to 
$\langle \Upsilon \vert O_1(^3S_1)
\vert \Upsilon \rangle$.

The use of the asymptotic form 
for $f(z)$ may introduce some errors,
because the scale 
$\mu$ here is actually $m_b$, not $\mu\to\infty$. However,
with a large $m_b$ one may expect 
that it can provide a good order of magnitude estimate
with the asymptotic form. 
We will use Eq. (\ref{R}) later for
our numerical discussions.

We note that at this stage the 
state $G_s$ can be any particle with the
same quantum number
as $G^{a,+\mu} G^{a,+}_\mu$, i.e., $J^{PC}=0^{++}$.
The normalization constant $f_s$ depends on the
properties of the specific particle. 
In order to obtain the branching ratio
of $G_s$ as a scalar glueball, 
we have to evaluate $f_s$ with $G_s$
specified to be the scalar glueball. 
In the following we provide an estimate
based on QCD sum rule calculation.
\\

\noindent
{\bf 3. QCD sum rule 
calculation of the normalization constant}


The constant
$f_s$ has  dimension one in mass and is related to the
the product of local
operator:
\begin{equation}
\langle G_s(k)\vert G^{\mu\rho} {G^\nu}_\rho \vert 0\rangle
  = f_0 m_G^2 g^{\mu\nu} +f_s k^\mu k^\nu.
\end{equation}
The fact that the same $f_s$ appears in Eq.(9) 
and Eq.(12) can be checked by integrating
over $z$ on the both sides of Eq.(6).
\par
The basic idea of the QCD sum rule 
calculation for our estimate is to
consider the two point correlator\\
\begin{eqnarray}\label{cor}
\Pi_{\mu\nu,\mu^\prime\nu^\prime}(Q^2)&=&\int d^4x e^{iq\cdot
x}i\langle 0|T{G_{\mu\alpha}G_\nu^\alpha(x),
G_{\mu^\prime\beta}
G_{\nu^\prime}^\beta(0)}|0\rangle\nonumber\\
&=&T_{\mu\nu\mu^\prime\nu^\prime}\Pi_T(Q^2)+
+V_{\mu\nu\mu^\prime\nu^\prime}\Pi_V(Q^2)
+S^1_{\mu\nu\mu^\prime\nu^\prime}\Pi_{S1}(Q^2)\nonumber\\
&&+S^2_{\mu\nu\mu^\prime\nu^\prime}\Pi_{S2}(Q^2)
+S^{3}_{\mu\nu\mu^\prime\nu^\prime}\Pi_{S3}(Q^2),
\end{eqnarray}
for a region of $Q$ in which 
one can incorporate the asymptotic
freedom property of QCD via the 
operator product expansion (OPE),
and then relate it to the hadronic matrix elements via the
dispersion relation. 
The tensors in Eq. (\ref{cor}) are defined as
\begin{eqnarray}
T_{\mu\nu\mu^\prime\nu^\prime}
&=&\displaystyle{g^t_{\mu\mu^\prime}
g^t_{\nu\nu^\prime}+g^t_{\mu\nu^\prime}g^t_{\nu\mu^\prime}
-\frac{2}{3}g^t_{\mu\nu}g^t_{\mu^\prime\nu^\prime}}
\nonumber\\
V_{\mu\nu\mu^\prime\nu^\prime}&=& g^t_{\mu\mu^\prime}q_\nu
q_{\nu^\prime} +g^t_{\nu\nu^\prime}q_\mu q_{\mu^\prime}
+g^t_{\mu\nu^\prime}q_\nu q_{\mu^\prime}
+g^t_{\nu\mu^\prime}q_\mu q_{\nu^\prime}
\nonumber\\
S^1_{\mu\nu\mu^\prime\nu^\prime}&=&
g^t_{\mu\nu}g^t_{\mu^\prime\nu^\prime},\;\;
S^2_{\mu\nu\mu^\prime\nu^\prime}=
g^t_{\mu\nu}q_{\mu^\prime}q_{\nu^\prime}
+g^t_{\mu^\prime\nu^\prime}q_{\mu}q_{\nu},\;\;
S^3_{\mu\nu\mu^\prime\nu^\prime}=
q_{\mu}q_{\nu}q_{\mu^\prime}q_{\nu^\prime},
\end{eqnarray}
where
$g^t_{\mu\nu} = g_{\mu\nu} -q_\mu q_\nu/q^2$.
The corresponding terms $\Pi_T(Q^2)$, $\Pi_V(Q^2)$,
$\Pi_{S1}(Q^2)$, $\Pi_{S2}(Q^2)$ 
and $\Pi_{S3}(Q^2)$  are from the
contributions of $2^{++}$, 
$1^{-+}$ and $0^{++}$ states respectively.

In a deep Euclidean 
region $Q^2=-q^2>>\Lambda_{QCD}$, they can be
expanded as
\begin{eqnarray}
\Pi_{i}(Q^2)=C^0_i(Q^2)I+C^1_i(Q^2)\alpha_s \langle
G_{\mu\nu}G^{\mu\nu}\rangle +C^2_i(Q^2)\langle
g_sf^{abc}G^{a~\mu}_{~~~\alpha}G^{b~\alpha}_{~~~\beta}
G^{c~\beta}_{~~~\mu}\rangle
+\cdots,
\end{eqnarray}
where $C^j_i$ are 
Wilson coefficients which need to be determined
later.

On the other hand, the correlator 
in Eq.(\ref{cor}) can be saturated
by all possible resonances and continuum. We have
\begin{equation}\label{phys}
Im\Pi_{\mu\nu,\mu^\prime\nu^\prime}(Q^2) 
=\displaystyle{\sum_R
\langle 0|G_{\mu\alpha}G_\nu^\alpha |R\rangle\langle R|
G_{\mu^\prime\beta}G_{\nu^\prime}^\beta|0\rangle }
\pi\delta(Q^2+m^2_R)+continuum,
\end{equation}
where the sum on $R$ are for all possible resonances.
The term $\langle \vert G_{\mu\alpha} G^\alpha_\nu \vert R
\rangle \langle R\vert G _{\mu'\beta}G^\beta_{\nu'}
\vert 0\rangle$ in
the above equation contains the 
information of $f_0$ and $f_s$ when
$R$ is the scalar glueball. 
The $T$ and $V$ tensors are not related to
$f_s$. They are irrelevant to our calculations. 
The functions $\Pi_{S1,S2,S3}$
contain linear combinations of $f_0$ and $f_s$.
QCD sum rule calculations for 
$\langle G_s\vert G^{\mu\nu}G_{\mu\nu}\vert
0 \rangle = (4f_0+ f_s)m^2_G$ 
has been carried out before\cite{8}.
Therefore if one of the 
$\Pi_{S1,S2,S3}$ is know, one can obtain $f_s$.
From eq. (10) and the tensor structure of eq. (11),
we find that $\Pi_{S3}$ is proportional to $(f_0+f_s)^2$.
Therefore the study of
$\Pi_{S3}$ is 
sufficient for our purpose of determining $f_s$.
$\Pi_{S1,2}$ also contain 
information about $f_0$ and $f_s$. However
the non-perturbative 
contributions for them begin at the level
of dimension-8 operators.
The results obtained are 
not as reliable as the one from $\Pi_{S3}$
which has a lower dimension. 
We now concentrate on $\Pi_{S3}$.

There may be several bound states 
with the same quantum numbers to include 
in the QCD sum rule calculation, such as a 
pure scalar glueball, 
quark bound states and higher excited states. The 
contributions from higher
excited states are suppressed 
upon the use of Borel transformation which 
is discussed in the below. For the quark bound states,
OZI rule implies that the conversion of bound quark state 
into a scalar glueball is suppressed compared with
the conversion of two gluon into a 
scalar glueball\cite{pen}, perturbatively 
suppressed by powers in $\alpha_s$. 
If this is indeed true, the corresponding 
$f_s$ parameters for quark bound states 
will be smaller than pure glueball
state. We will work with this approximation 
in the following discussions.
To be consistent with our previous expansion,
we again work to order $\alpha_s$.
To this order, 
using the method in Ref. \cite{10}, we find
\begin{eqnarray}\label{ope}
\Pi_{S3}(Q^2)&=&\displaystyle{\frac{1}{8\pi^2}\ln
\frac{\mu^2}{Q^2}+ \frac{1}{2Q^4}(\langle
G_{\mu\nu}G^{\mu\nu}\rangle +\frac{2g_s}{Q^6}\langle
f^{abc}G^{a~\mu}_{~~~\alpha}G^{b~\alpha}_{~~~\beta}
G^{c~\beta}_{~~~\mu}\rangle)}.
\end{eqnarray}

The correlator in Eq. (\ref{ope}) 
obtained by using OPE is  related to
Eq. (\ref{phys}) via the standard dispersion relation
\begin{equation}\label{dis}
\Pi_{\mu\nu,\mu^\prime\nu^\prime}(Q^2)
=\displaystyle{\frac{1}{\pi}\int^\infty_0
ds\frac{Im\Pi_{\mu\nu,\mu^\prime\nu^\prime}(-s)}{s+Q^2}}.
\end{equation}
In practice one 
may only include ground states in the calculation.
In order to reduce uncertainty due to higher
excited states and also continuum states,
we apply Borel transformation and obtain

\begin{equation}
\hat B\Pi_{S3}(Q^2)
=\displaystyle{\frac{1}{M^2}\int^{s_0}_0 ds
e^{-s/M^2}\rho_{S3}(s)},
\end{equation}
where $\rho_{S3}(s) = (1/\pi) Im\Pi_{S3}(-s)$, and

\begin{equation}\label{bor}
\hat B\Pi(Q^2)=\lim_{
Q^2,n\rightarrow\infty}
\displaystyle{\frac{1}{(n-1)!}(Q^2)^n
(-\frac{d}{dQ^2})^n
\Pi(Q^2)},
\end{equation}
Here one also needs to have the 
limit $Q^2/n = M^2 = constant$.

In our numerical calculation
we have varied $s_0$ in the range of  
$3-6GeV^2$, and found that the
uncertainty  is around 10 percent. 
The parameters determined are reasonably stable.

We obtain the range for $f_s$ as
\begin{equation}
f_s=(100\sim 130){\rm MeV},
\end{equation}
with
$f_0 = 190$ MeV and 
$f_s = 100$ MeV for $m_{0^{++}} = 1.5$ GeV, and
$f_0 = 130$ MeV
and $f_s = 130$ for $m_{0^{++}} = 1.7$ GeV.
In obtaining the above result, 
we have re-evaluated $f_0$ also using the same
parameters.
The input parameters used are\cite{9}
$\alpha_s(\mu)=
4\pi/9\ln(\mu^2/\Lambda^2_{QCD})$,
$\Lambda_{QCD}=0.25GeV$,  $\mu=M$,
$\langle\alpha_sG_{\mu\nu}G^{\mu\nu}\rangle
=0.06\pm0.02GeV^4$, and
$g_s\langle f^{abc}G^{a~\mu}_{~~~\alpha} 
G^{b~\alpha}_{~~~\beta}
G^{c~\beta}_{~~~\mu}
\rangle=0.27GeV^2\langle\alpha_sG^{\mu\nu} 
G_{\mu\nu}\rangle$.

For consistency, 
we also calculated the glueball masses. We find that
for the  $0^{++}$ state the mass 
is $1.5\sim 1.7$ GeV, and for $2^{++}$ the mass
is $2.0\sim 2.2$ GeV. 
These are in agreement with other calculations\cite{8}.

If the scalar glueball 
is a pure one, using the above results we obtain
the branching ratio for 
$\Upsilon \to \gamma G_s$ to be in the range

\begin{eqnarray}
Br(\Upsilon \to \gamma G_s) = (1 \sim 2)\times 10^{-3},
\end{eqnarray}
with a larger branching ratio for a 
larger glueball mass up to 1.7 GeV.
Here we have used $\alpha_s = 0.18$ which is the typical
value for $\alpha_s$ in the energy range of the decay.
We obtain a 
large branching ratio for $\Upsilon \to \gamma
G_s$. We would like to 
point out that considering several uncertainties,
the assumptions of factorization 
and single pure glueball state in the
QCD sum rule calculation, the above 
numbers should be used as an order of magnitude estimate.
\\

\noindent
{\bf 4. Discussions of phenomenological implications}

Experimental measurement of 
$\Upsilon \to \gamma G_s$ may be non-trivial.
One has to rely on the decay products of 
glueballs. There are several ways
the glueball can decay with 
reasonably large branching ratios,
$G_s \to K \bar K$ or 
$G_s$ to multi-pions. As mentioned earlier that there
are several candidates for 
scalar glueball, the $f(1370)$, $f_0(1500)$ and $f_0(1710)$.
Decays of $\Upsilon \to \gamma f_0(i) \to
\gamma (K\bar K \mbox{or multi-pions})$ 
can provide important information.

Experimentally there is only an upper bound\cite{4} 
of 
$Br(\Upsilon \to \gamma f_0(1710) \to \gamma K \bar K)
<2.6\times 10^{-4}$ at 90\% C.L..
If $f_0(1710)$ is a pure glueball,
experimental measurement\cite{4} of
$Br(f_0(1710) \to K \bar K) 
= 0.38^{+0.09}_{-0.19}$\cite{4}
would imply 
$Br(\Upsilon \to \gamma f_0(1710) \to \gamma K \bar K)$
to be in the range of 
$ (0.4 \sim 1.0)\times 10^{-3}$ which seems to indicate that
$f_0(1710)$ may not be a pure glueball. At present it 
can not rule out the possibility that one of the
$f_0(1370)$ or $f_0(1500)$ being a pure glueball state.
Data also allow certain mixing among glueball state and other
quark bound states.

Theoretical calculation of 
the mixings among glueball and quark bound
states is a very difficult task. 
There is no reliable theoretical
calculation. Lattice calculations may 
eventually give accurate predictions for
the mixing parameters. At present there 
are some phenomenological studies of
glueball mixings. We now study some 
implications of the branching ratio
for the radiative decay
of a $\Upsilon$ into a pure scalar 
glueball obtained in the previous section
on a mixing pattern suggested in Ref. \cite{11}.

An analysis combining other 
experimental data in Ref.\cite{11}
showed that the three $0^{++}$ states 
$f_0(1370)$, $f_0(1500)$ and $f_0(1710)$
all contain substantial glueball content. 
Ref.\cite{11} obtained a mixing
matrix of physical states in terms of
pure glueball and 
other quark bound states to be\cite{11}

\begin{eqnarray}
\begin{array}{llll}
              & 
f_{i1}^{G_s} & f_{i2}^{S} & f_{3i}^{(N)}\\
f_0(1710)     & 
0.39 \pm 0.03 & 0.91\pm 0.02& 0.15\pm 0.02\\
f_0(1500)     & 
-0.65\pm 0.04& 0.33\pm 0.04& -0.70\pm 0.07\\
f_0(1370)     & 
-0.69\pm 0.07& 0.15\pm0.01&0.70\pm 0.07
\end{array}
\label{mix}
\end{eqnarray}
where the states $G_s$, $S = |s\bar s\rangle$ and 
$N = |u\bar u +d\bar d\rangle/\sqrt{2}$
are the pure glueball and quark bound states.
$f_{i1}^{G_s}$ indicate the amplitude 
of glueball $G_s$ in the three
physical $f_0(i)$ states.

Because of the mixing, when applying our calculations to
radiative decay of 
$\Upsilon$ into a physical state which is not
a purely gluonic state the parameters 
will be modified. If the mixing parameter
is known one can obtain the $R_s$ ratios for
$\Upsilon \to \gamma f_0(1370,1500,1710)$ as
\begin{equation}
R_s(\Upsilon\to \gamma f(i))
  = \frac{25\pi\alpha_s^2}{3\alpha} 
\cdot\frac{\vert f_s\vert^2}{m_b^2}
\vert f^{(G_s)}_{i1}\vert^2.
\end{equation}
$\Upsilon \to \gamma f(i)$ may also result from
$\Upsilon $ decays into a $\gamma$ and 
$S, N$ quark bound states. However,
these processes are suppressed by $\alpha_s^2$.

Using the mixing amplitudes in Eq.(\ref{mix}),
one obtains the branching ratios of,
$\Upsilon \to \gamma 
f_0(1370, 1500, 1710)$ to be in the ranges,
$(4.8\sim 9.6, 4.2\sim 8.4, 1.5\sim 3.0)\times 10^{-4}$.
Combining the branching ratios
of $f_0(1370,1500,1710) \to K \bar K (\pi\pi)) =$
$(0.38^{+0.09}_{-0.19} (0.039^{+0.002}_{-0.024}), 
0.044^{+0.021}_{-0.021}
(0.454^{+0.104}_{-0.104}), 
0.35^{+0.13}_{-0.13}(0.26^{+0.09}_{-0.09}))$\cite{4}, we
obtain:

\begin{eqnarray}
\begin{array}{ll}
Br(\Upsilon \to \gamma f_0(1710) \to \gamma K \bar K)
\approx 0.6\sim 1.2; &
Br(\Upsilon \to \gamma f_0(1710) \to \gamma \pi\pi) 
\approx 0.06 \sim 0.12; \nonumber\\
Br(\Upsilon \to \gamma f_0(1500) \to \gamma K \bar K)
\approx 0.2\sim 0.4; &
Br(\Upsilon \to \gamma f_0(1500) \to \gamma \pi\pi) 
\approx 1.9\sim 3.8; \nonumber\\
Br(\Upsilon \to \gamma f_0(1370) \to \gamma K \bar K)
\approx 1.7\sim 3.4; &
Br(\Upsilon \to \gamma f_0(1370) \to \gamma \pi\pi) 
\approx 1.2\sim 2.4.
\end{array}
\end{eqnarray}
In the above the branching ratios are 
in unit $10^{-4}$. 
The branching ratios predicted above can provide further 
test for QCD factorization. 
Future experimental data from CLEO III will provide us with 
important information.

To summarize, we have estimated the 
branching ratio of $\Upsilon\to\gamma+G_s$
with $G_s$ as a glueball. Our result 
shows that $f_0(1710)$ may not be consistent with the assumption that 
it is a pure glueball, but can not rule out the possibility 
that one of the $f(1370)$, 
and $f_0(1500)$ being a pure glueball state.
We also predicted several $\Upsilon \to \gamma KK(\pi\pi)$ branching ratios
using a phenomenological glueball mixing pattern 
which can provide further tests for QCD factorization calculations and glueball
mixing.
To have a better understanding of 
the situation, we have to rely on future
improved experimental data. 
Fortunately CLEO-III will provide us with more
data in the near future. 
We have a good chance to understand
the properties of scalar glueball. 
We strongly encourage our experimental
colleagues to carry out the study of 
radiative decay of $\Upsilon$ into
a scalar glueball.

\vskip20pt
\noindent
{\bf\large Acknowledgments}

The work of X.G.H.
was supported in part by 
National Science Council under grants NSC
89-2112-M-002-058.
The work of H.Y.J. and J.P.M is 
supported  by National Nature
Science Foundation.
\vfil\eject


\begin{thebibliography}{99}


\bibitem{2} C.J. Morningstar and M. Peardon,
Phys. Rev. {\bf D56}, (1997) 4043,
C. Liu, Nucl. Phys. Proc. Suppl.
{\bf 94}, (2001) 255, 
C. Liu, Chin. Phys. Lett. {\bf 18}, (2001) 187

\bibitem{3} A. Anastassov et al.(CLEO Collaboration),
Phys. Rev. Lett. {\bf 82}, (1999) 286;
S. Richichi et al.(CLEO Collaboration),
Phys. Rev. Lett. {\bf 87}, (2001) 141801;
CLEO Collaboration, {\it Further Experimental
Studies of Two-Body Radiative $\Upsilon$ Decays},
hep-ex/0201005.

\bibitem{4} Particle Data Group, 
Euro. Phys. Jur. {\bf C15}, (2000) 1.

\bibitem{1} L. Gibbons, 
{\it The proposed CLEO-C program and
R measurement prospects},
hep-ex/0107079

\bibitem{5} G.T. Bodwin, E. Braaten, and G.P. Lepage,
Phys. Rev. {\bf D51}, (1995) 1125; {\it ibid.} {\bf 55},
(1997) 5853(E).

\bibitem{6} J.P. Ma, Nucl.Phys. {\bf B605}, (2001) 625;
Erratum-ibid. {\bf B611}, (2001) 523.

\bibitem{12} J. P. Ma, hep-ph/0202256.

\bibitem{7}
M. A. Shifman, A. I. Vainshtein 
and V. I. Zakharov, Nucl. Phys.
{\bf B147},
(1979) 385.

\bibitem{8}
T. Huang H.Y. Jin and A.L. Zhang, 
Phys. Rev. {\bf D59}, (1999) 4026;
S. Narison, Nucl. Phys. {\bf B502}, (1998) 312.

\bibitem{9} L.J. Reinders, 
H. Rubinstein and S. Yazaki, Phys. Rep.
{\bf 127}, (1985) 1; 
S. Narison, Phys. Lett. {\bf B387}, (1996) 162.

\bibitem{10}
V.A. Novikov, M. A. Shifman, A. I. Vainshtein and
V. I. Zakharov, Nucl. Phys. {\bf B174}, (1980) 378.

\bibitem{pen} M. Pennington, e-print hep-ph/9811276.

\bibitem{11} F. Close and A. Kirk, 
Eur.Phys.J. {\bf C21}, (2001) 531


\end{thebibliography}
\end{document}